\documentclass[12pt,preprint]{aastex}

\begin{document}

\title{51 Eri and GJ 3305: A $10-15$ Myr old binary star system at 30 parsecs}

\author{E.\ D.\ Feigelson$^1$, W.\ A.\ Lawson$^2$, M.\ Stark$^{1,3}$,L.\ Townsley$^1$, G.\ P.\ Garmire$^1$}

\altaffiltext{1}{Department of Astronomy \& 
Astrophysics,Pennsylvania State University, University Park PA 
16802.  Email: edf@astro.psu.edu}

\altaffiltext{2}{School of Physical, Environmental \& Mathematical 
Sciences, University of New South Wales, Australian Defence Force 
Academy, Canberra ACT 2600, Australia}

\altaffiltext{3}{Department of Physics \& Astronomy, University of 
Wyoming, Dept. 3905, 1000 East University Avenue, Laramie WY 82071}

\slugcomment{Resubmitted to the Astronomical Journal, Nov. 16, 2005}

\begin{abstract}

Following the suggestion of Zuckerman et al.\ (2001, ApJ, 562, L87), 
we consider the evidence that 51 Eri (spectral type F0) and GJ 3305 
(M0), historically classified as unrelated main sequence stars in 
the solar neighborhood, are instead a wide physical binary system 
and members of the young $\beta$ Pic moving group (BPMG).  The BPMG 
is the nearest ($d \la 50$ pc) of several groups of young stars with 
ages around 10 Myr that are kinematically convergent with the 
Oph-Sco-Cen Association (OSCA), the nearest OB star association. 
Combining SAAO optical photometry, Hobby-Eberly Telescope 
high-resolution spectroscopy, $Chandra$ X-ray data, and UCAC2 
catalog kinematics, we confirm with high confidence that the system 
is indeed extremely young.  GJ 3305 itself exhibits very strong 
magnetic activity but has rapidly depleted most of its lithium.  The 
51 Eri/GJ 3305 system is the westernmost known member of the OSCA, 
lying 110 pc from the main subgroups.  The system is similar to the 
BPMG wide binary HD 172555/CD $-64^\circ$1208 and the HD 104237 
quintet, suggesting that dynamically fragile multiple systems can 
survive the turbulent environments of their natal giant molecular 
cloud complexes, while still being imparted high dispersion 
velocities.  Nearby young systems such as these are excellent 
targets for evolved circumstellar disk and planetary studies, having 
stellar ages comparable to that of the late phases of planet 
formation.

\end{abstract}


\section{Introduction \label{intro.sec}}

\subsection{Young stars near Earth are mostly OSCA outliers}

Observations revealing the transition from gaseous circumstellar 
disks to protoplanets are not limited by the capabilities of 
telescopes but by the available samples of nearby young stars.  
Pre-main sequence (PMS) evolutionary models indicate that a 10 Myr 
old protoplanet with mass $\geq 10$ M$_J$ should exhibit $I<20$ at a 
distance of 30 pc \citep{Baraffe02}.  The principal difficulties in 
finding stars with dissipating disks or protoplanets are that no 
active star forming regions lie within $\sim 150$ pc and that older 
PMS stars drift far from their natal clouds where they are difficult 
to distinguish from the Galactic field population.

Most known nearby older PMS stars are kinematically linked to the 
populous Ophiuchus-Scorpius-Centaurus Association (OSCA) of PMS 
stars with its main concentrations at mean distances of $118-145$ pc 
\citep{Blaauw91, deZeeuw99}.   Ages range from 17 Myr in the 
Lower-Centaurus-Crux (LCC) subgroup \citep{Mamajek02} to $<1$ Myr in 
the Ophiuchus cluster where stars are still forming today.  Several 
sparse populations distributed over the southern sky are 
kinematically convergent with the rapidly-moving OSCA and exhibit 
circumstellar disks, lithium excesses, or other indicators of 
stellar youth. These include the TW Hya Association (TWA), $\beta$ 
Pic moving group (BPMG), $\eta$ Cha cluster, $\epsilon$ Cha group, 
and several `isolated' Herbig AeBe stars with low mass companions.  
Their positions in the sky are shown in Figure \ref{OSCA.fig} 
\citep[a similar figure with proper motions vectors and references 
appears in][]{Feigelson03}.  These stars are distributed over a 
large region around the rich OSCA concentrations, likely propelled 
by supersonic motions of gaseous eddies in the OSCA's turbulent 
giant molecular cloud \citep{Feigelson96}.  It is quite likely that 
other stellar systems also lie away from the main OSCA subgroups.  
The BPMG in particular has members very close to the Sun. The 
circumstellar disk of its most massive member, $\beta$ Pic ($d = 19$ 
pc), has been subject to very intensive study \citep{Lagrange00} and 
a disk has recently been reported around the BPMG member AU Mic 
\citep[$d = 10$ pc;][]{Kalas04}.

We discuss here two stars proposed by \citet{Zuckerman01} to be both 
a physical binary and nearby members of the $\sim 12$ Myr old BPMG:  
the $V = 5.2$ F0 star 51 Eri and the $V=10.6$ M0.5 star GJ 3305 
separated by 66\arcsec.  This suggestion is in conflict with 
previous studies of these stars which have considered them to be 
unrelated older main sequence stars at $d=31$ pc and $d=15$ pc, 
respectively (\S \ref{history.sec}).  We present new X-ray and 
optical observations (\S \ref{obs.sec}), and consider in detail the 
kinematic, spectroscopic, photometric and magnetic activity 
properties of these stars (\S \ref{evol.sec}). We conclude that they 
indeed are outlying members of the OSCA.  51 Eri and GJ 3305 are 
thus among the nearest $10-15$ Myr old stars and represent one of 
the best nearby young stellar systems for study of planet formation. 
Their protoplanetary disks have largely dissipated and protoplanets 
might be detectable.

\subsection{The misclassification of GJ 3305 \label{history.sec}}

The literature on 51 Eri and nearby associated stars has been 
confused.  Figure \ref{images.fig}$a$ shows the neighborhood from 
the 2MASS $J$-band survey, and Table \ref{vicinity.tab} lists the 
labeled stars.  Many of these stars are missing from star catalogs 
due to saturation and diffracted light from the $V = 5.2$ 51 Eri, 
especially those based on all-sky Schmidt telescope photographic 
surveys. 51 Eri itself has a {\it Hipparcos\,} parallax measurement 
which, when combined with photometry and spectroscopy, confirms that 
it is a F0 star positioned on, or very near, the main sequence at a 
distance of $29.8 \pm 0.8$ pc (for convenience, we call this 30 pc 
throughout this paper). From Burnham's report in 1916, it was 
considered a double star with star 7 as its companion.  But this is 
incorrect: kinematic measurements from the recent UCAC2 survey 
\citep[Table \ref{vicinity.tab};][]{Zacharias03} established that 
star 7 does not share the proper motions of 51 Eri and must be an 
unrelated star.

GJ 3305 has been classified, incorrectly in our view, as a disk 
dwarf with a photometric distance of $\approx 15$ pc based on the 
assumption that the star resides on the main sequence.  Propagation 
of this assumption has lead to the inclusion of GJ 3305 in other 
catalogs and surveys of nearby stars, e.g. the Nearby Stars 
catalog\footnote{GJ 3305 does not appear in their published catalog 
but was added by the SIMBAD staff as a NN `new neighbour' addition 
to the GJ catalog.} of \citet{Gliese95} and the Palomar/MSU nearby 
star spectroscopic survey of \citet{Gizis02}. GJ 3305 is cataloged 
as a chromospherically-active flare star, presumably on account of 
its H$\alpha$ emission, and more-recently of detection of strong 
X-ray variability by \citet{Fuhrmeister03}.  Their calculation of 
the total energy released during an X-ray flare detected during the 
{\it ROSAT\,} All-Sky Survey of $3.4 \times 10^{34}$ erg is derived 
assuming a distance of 15.2 pc, with this value rising by a factor 
$\approx 4$ if GJ 3305 instead resides at 30 pc as we argue in this 
paper.  Across the characteristic timescale of the flare (we adopt 
0.2 days) the luminosity of the flare was $L_{X} \approx 8 \times 
10^{30}$ erg\,s$^{-1}$.  For a spectral type of M0.5 and $V = 10.6$ 
for GJ 3305 at 30 pc, and adopting the dwarf temperature and 
bolometric correction sequences of \citet{Kenyon95} as is 
appropriate for `older' PMS stars \citep{Mamajek02, Lyo04}, we 
derive a stellar luminosity $L_{bol} = 0.14\, L_{\odot}$ and find 
the ratio of X-ray flux to bolometric flux to be $\log L_X/L_{bol} 
\approx -1.8$ during the flare.  Our {\it Chandra\,} observation 
indicates a lower quiescent level for GJ 3305 of $\log L_{X}/L_{bol} 
\approx -3.5$, a value similar to the `saturation' level of $\log 
L_{X}/L_{bol} \sim -3$ observed for magnetically-active late-type 
PMS stars.

\section{Observations \label{obs.sec}}

\subsection{Chandra X-Ray Observatory  \label{chandra.sec}}

A $16\arcmin \times 16\arcmin$ region around 51 Eri was observed for 
3.15 ks with the Advanced CCD Imaging Spectometer (ACIS) detector on 
board the {\it Chandra\,} {\it X-Ray\,} {\it Observatory\,}.  The 
instrument and detector are described by \citet{Weisskopf02}.  This 
snapshot observation took place on 2003 November 20.  Data analysis 
followed procedures described in \citet{Townsley00}, 
\citet{Townsley03} and \citet{Getman05} using the IDL- and 
CIAO-based script {\it acis\_extract\,} package\footnote{Description 
and code for {\it acis\_extract\,} are available at \\ 
\url{http://www.astro.psu.edu/xray/docs/TARA/ae\_users\_guide.html}.}.  
One processing step, the removal of $\pm 0.25$\arcsec\/ 
randomization in event location, was omitted due to the short 
exposure treated here.  Figure \ref{images.fig}$b$ shows the central 
$6\arcmin \times 6\arcmin$\/ of the resulting ACIS image after 
correction for charge transfer inefficiency and data selection 
steps.  A spatial offset of $0.7$\arcsec\/ was applied to align the 
sources associated with 51 Eri and GJ 3305 to the {\it Hipparcos\,} 
reference frame.

Thirty-one candidate X-ray sources were located using a 
wavelet-based detection algorithm \citep{Freeman02}.  Events for 
each source were extracted using {\it acis\_extract}. Here, events 
are extracted in a small region around each source containing 95\% 
of the enclosed energy derived from the point spread function of the 
telescope at that position, and a local background is defined from a 
nearby source-free region. The extraction of GJ 3305 required a 
special annular region because the central pixels were subject to 
photon pileup, a saturation of the detector occurring when more than 
one photon arrives in a pixel during the 3.2 s between CCD readouts.  
We extracted 30\% of the incident photons arriving between the 60\% 
and 90\% enclosed energy circles, and adjusted the effective area 
(CIAO's {\it arf\,} file) accordingly.

Most of these sources are extragalactic. Five were detected in a 
{\it ROSAT\,} pointed observation (WGACAT 0437.2-0210, 
0437.3-0222,0437.5-0236, 0437.6-0231 and 0437.8-0230) but have not 
been studied. Four have $R \simeq 19-20$ counterparts on the 
Digitized Sky Survey and one is associated with the $V \simeq 14.5$ 
S0 galaxy UGC 03105 at redshift $z = 0.03$.  The remaining optical 
counterparts are too faint to appear in all-sky surveys.  These 
sources are typical of the extragalactic population seen in 
$Chandra$ images which often are active galactic nuclei with 
$0.1<z<1$ \citep{Brandt05}.

Table \ref{xray.tab} provides detailed results for the two stars of 
interest here.  CXOU is the official designation of the $Chandra$ 
X-ray sources.  The positions have a precision of $\pm 0.2$\arcsec. 
{\it Extracted counts} gives the number of extracted counts in the 
total $0.5-8.0$ keV $Chandra$ band, while {\it Soft Counts} gives 
the subset of counts in the $0.5-2.0$ keV band.  Spectral analysis 
of 51 Eri and GJ 3305 is based on optically-thin plasma models with 
solar elemental abundances where the plasma energies $kT$ are 
obtained by least-squares fit of the photon energy distribution 
using the {\it XSPEC} software package \citep{Arnaud96}. No 
interstellar absorption is seen in the spectra. The spectrum and 
best-fit model for GJ 3305 is shown in Figure \ref{xspec.fig}.  It 
requires a two-temperature plasma which is common in ACIS spectra of 
magnetically active late-type stars. Fluxes are found by integrating 
the best-fit spectrum over the $0.5-8$ keV band,and luminosities are 
calculated assuming the {\it Hipparcos\,} distance to 51 Eri of 30 
pc.

\subsection{Hobby-Eberly Telescope spectroscopy  \label{het.sec}}

High-resolution spectra of 51 Eri and GJ 3305 were taken on three 
sequential nights, 2003 December $13-15$ with the High Resolution 
Spectrograph \citep{Tull98} on the 9.2-meter Hobby-Eberly Telescope 
\citep{Ramsey98}.  The HRS uses an echelle mosaic with 
cross-dispersing gratings imaging onto a mosaic of two thinned 2k 
$\times$ 4k CCD detectors with 15 $\mu$m pixels. The instrument lies 
in a stationary, climate-controlled room and is fiber-coupled to the 
primary focus of the telescope.  The chosen 600 gr mm$^{-1}$ grating 
with 2\arcsec\/ fibers gives resolving power $R = 60,000$ over the 
interval $\sim 5300-7250$ \AA. We obtained one 30~sec exposure of 
51~Eri, and two 900~sec exposures of GJ 3305 each night, along with 
instrument calibration exposures.

The spectra were processed, extracted, and wavelength calibrated 
with standard IRAF routines.  A difficulty arose because the 
instrument continuum calibration lamp for flat-fielding sometimes 
exhibits lithium Li I $\lambda$6708\AA\, (sodium Na D) emission 
lines with 18\% (12\%) amplitude that coincide with stellar lines of 
interest.  We replaced 2 \AA\, around each of these lines with a 
linear interpolation of the surrounding flat lamp spectrum; $\approx 
2$\% systematic residual structure is expected from this procedure.  
For the GJ 3305 spectra, background spectra from two fibers pointed 
at blank sky were subtracted and cosmic rays were rejected. These 
steps were unnecessary for 51 Eri owing to the brightness and 
brevity of the exposures.  Spectra from the three nights were 
combined and the continua of the wavelength-calibrated star spectra 
were removed.  No radial velocity correction or calibration was 
made.  No night-to-night variation in any spectral feature was seen.

Figure \ref{het.fig} shows the resulting spectra around the Na 
D,H$\alpha$ and Li I $\lambda$6708\AA\, lines.  The signal-to-noise 
ratio at H$\alpha$ and Li I $\lambda$6708\AA\, is S/N = 230:1 for GJ 
3305 and 180:1 for 51 Eri.  The strength of the lithium absorption 
line in GJ 3305 is of greatest interest in our discussion below. 
Figure \ref{het.fig}$c$ shows the spectra with a relatively low 
global continuum fit which gives a Li I $\lambda$6708\AA\, 
equivalent width $EW = 0.09$ \AA.  A continuum fit that passes near 
the peaks of local fluctuations, which takes into account our high 
$S/N$ and the likely presence of uncatalogued absorption lines, 
gives $EW = 0.14$ \AA. (Jeffries et al.\ 2003 discuss the 
sensitivities of reported lithium line strengths to the choice of 
continuum levels.)  A faint absorption line with $EW = 0.03$ \AA\, 
appears in the 51 Eri spectrum; we believe this is the residual from 
the interpolated continuum lamp calibration. The true GJ 3305 line 
strength may thus be 0.03 \AA\, lower than measured here.  
Altogether, we adopt $EW = 0.12 \pm 0.03$ for the Li I 
$\lambda$6708\AA\, line of GJ 3305.

The weak H$\alpha$ profile of GJ 3305 with roughly 1 \AA\, $EW$ and 
little variability over 3 nights of observation is commensurate with 
GJ 3305 being a weakly-active, non-accreting PMS star.  It has a 
spectroscopic analog in the several-Myr younger star RECX 10 in the 
$\eta$ Cha cluster, with a spectral type of M0.3 determined from 
precise spectrophotometric study \citep{Lyo04}.  Both stars have 
weak, narrow H$\alpha$ profiles of 1 \AA\, $EW$ and velocity width 
(measured at the 10\% height of the emission line profile above the 
surrounding continuum) $v_{10} = 110$ km\,s$^{-1}$ indicative of a 
line of chromospheric origin, with no evidence for wing-broadening 
that might indicate on-going low levels of disk accretion 
\citep{Lawson04}.  At echelle resolution, the H$\alpha$ profiles of 
both stars show weak self-absorption signatures.

In the spectral regions of interest here, the principal difference 
between GJ 3305 and RECX 10 is the strength of the Li I 
$\lambda$6708\AA\, line; $EW =  0.12$ \AA\, for GJ 3305 (see the 
discussion above) and $EW = 0.6$ \AA\, for RECX 10 (measured from 
unpublished echelle spectra; Mamajek et al.\ 1999 had earlier 
obtained a similar result of $EW = 0.5$ \AA\, from medium-resolution 
spectra).  Our measurement of low lithium $EW$ in GJ 3305 reinforces 
the evidence that significant lithium depletion can be present in 
$10-15$ Myr stars (\S 6).  \citet{Zuckerman01} used lithium 
measurements as evidence that the BPMG has an age intermediate 
between that of the TWA ($\sim 10$ Myr) and Tucana-Horologium ($\sim 
30$ Myr) group stars.  Adding our lithium measurement for GJ 3305 to 
those for 3 late-type BPMG candidate members with spectral types 
ranging from K0 to M1 measured by \citet[][see their Table 
2]{Zuckerman01} gives an average lithium $EW = 0.26$ \AA.  GJ 3305 
has the lowest $EW$ by a factor of $\approx 2$ compared to the other 
three stars, and thus signals that depletion has been unusually 
efficient in this star.  This average $EW$ for BPMG candidate 
members compares to average values of $\approx 0.5$ \AA\, for 
late-type members of both the $\sim 10$ Myr-old TWA \citep{Webb99} 
and $\eta$ Cha star cluster \citep{Mamajek99}.

\subsection{South African Astronomical Observatory photometry}

We have obtained a tentative measure of the rotation period of GJ 
3305 by searching for optical photometric variations that are 
modulated by the periodic rotation of starspots.  A series of 
observations was made in the {\it BVRI\,} photometric bands over 6 
consecutive nights from 2004 December $8-13$ with the 1-m telescope 
and a 1k $\times$ 1k CCD detector at the South African Astronomical 
Observatory (SAAO).  Four sets of observations were obtained on most 
nights, to reduce the 1-day aliasing effects that can result from 
observations being made at a single observing site.  The 
observations and analysis of the photometry follows that described 
by \citet{Lawson01}.  In summary, differential measurements of the 
suspected variable star are produced by comparison with other stars 
in the CCD field that are assumed to be constant, and the resulting 
differential light curve is then subject to Fourier analysis.  
Unfortunately GJ 3305 resides in a sparse field towards the Galactic 
anti-center ($\ell, b = 199^{\circ}, -31^{\circ}$) resulting in few 
usable comparison stars within a radius of several arcmin.  With 
exposure times optimized for GJ 3305, only in $B$-band were there 
nearby field stars of comparable brightness to GJ 3305; in the other 
bands the usable comparison stars were a few magnitudes fainter than 
GJ 3305 resulting in unacceptable noise in the differential 
observations.  We therefore describe only the analysis of our 
$B$-band observations here.  

The $B$-band light curve shows variations of low amplitude over the 
duration of the observations.  Fourier analysis of the light curve 
recovered a periodicity at $f = 0.164$ d$^{-1}$, or $P = 6.1$ days 
(Figure \ref{phase.fig}). The recovered period is comparable to the 
length of the observing run, and so we treat the result with 
caution, but the data when phased to a period of $\approx 6$ days 
have less scatter than when they are phased to substantially 
different periods.  We therefore tentatively adopt a period of $P = 
6.0 \pm 0.5$ days for GJ 3305, and await the acquisition of 
photometry obtained across a longer timeline to confirm and improve 
this result.  In the upper panel of Figure \ref{phase.fig} we show 
our $B$-band photometry phased on a periodicity of $f = 0.164$ 
d$^{-1}$.  In the lower panel we show the amplitude spectra of the 
photometry (bold line), and that following pre-whitening of the 
dataset with the identified periodicity (thin line), across a 
frequency range of $0-2$ d$^{-1}$.  The collapse of the $f = 0.164$ 
d$^{-1}$ periodicity and its associated $(1-f)$ d$^{-1}$ and $(1+f)$ 
d$^{-1}$ aliases in the pre-whitened spectrum suggests the period of 
GJ 3305 is not far-removed from $P = 6$ days.  For a sine wave 
amplitude of 0.015 mag, and adopting a noise level of 0.003 mag over 
the $0-2$ d$^{-1}$ frequency interval, the periodicity has a S/N 
ratio of 5.  At higher frequencies both the original and 
pre-whitened spectra show noise at a level of $\approx 0.05$ mag, 
indicating that the variations seen within an individual night of 
data are the result of random photometric noise and not due to the 
presence of high frequency periodicities.

\section{Evolutionary status of 51 Eri and GJ 3305 \label{evol.sec}}

In this section we reevaluate the evolutionary state of these two 
stars based on various lines of argument.  It is worthwhile to 
recall that it is very difficult to determine the age of isolated 
AF-type stars on or near the main sequence from their spectral and 
photometric properties due to the convergence of isochrones in the 
HI diagram.  We note especially the extended debate concerning the 
age of the A3 star $\beta$ Pic. Once considered to be a post-main 
sequence star with age $t \simeq 200$ Myr or a main sequence star 
with age $t \simeq 100$ Myr, its kinematical association with PMS 
stars and the OSCA strongly argue for an age around $10-15$ Myr 
\citep{Barrado99, Mamajek01, Zuckerman01, Ortega02, Ortega04}. We 
synthesize several lines of evidence that together strongly argue 
that 51 Eri and GJ 3305 are also OSCA outliers with ages $10-15$ 
Myr.

\subsection{Astrometry and kinematics \label{kinem.sec}}

The UCAC2 proper motion of GJ 3305 ($\mu_\alpha,\mu_\delta$) = 
(46.1, -64.8) mas yr$^{-1}$ is nearly identical to the proper motion 
of $(\mu_\alpha,\mu_\delta) = (43.3, -64.3)$ mas\,yr$^{-1}$ found by 
{\it Hipparcos\,} for 51 Eri itself.  None of the other $\approx 
6000$ UCAC2 stars within an area of 10 deg$^{2}$ centered on the 
position of 51 Eri have a proper motion within 10 mas\,yr$^{-1}$ of 
that of 51 Eri.  We have already established that the 51 Eri motion 
is convergent with the OSCA $10-15$ Myr ago; see Figure 
\ref{OSCA.fig}.  The radial velocity of 51 Eri is $21 \pm 
2$km\,s$^{-1}$ \citep{Kharchenko04}.  Although we did not calibrate 
our HET spectra against radial velocity standards, differential 
velocity measurements of spectral features in 51 Eri and GJ 3305 
suggests both stars share the same radial velocity to within 2 
km\,s$^{-1}$.

\subsection{Magnetic activity \label{mag.sec}}

GJ 3305 has an X-ray to bolometric luminosity ratio $\log 
L_X/L_{bol} = -3.5$ which is near the saturation level of $\log 
L_X/L_{bol} \sim -3$ for late-type magnetically active PMS stars.  
This is far above the range of $\log L_X/L_{bol} = -5.2 \pm 0.5$ 
seen in old disk early-M type stars \citep{Fleming95}. A faint X-ray 
source with $L_X = 1.4 \times 10^{28}$ erg s$^{-1}$ and a soft 
spectrum coincides with the F0 primary, 51 Eri.  At 30 pc distance, 
this yields a low value of $\log L_X/L_{bol} = -6.2$ as is expected 
from the quiescent coronal gas in the atmosphere of an F0 star.

This high level of X-ray emission is a strong indicator for stellar 
youth.  We compare $\log L_X/L_{bol}$ of GJ 3305 with stars in other 
young stellar clusters and associations in Figure
\ref{Lx_Li_age.fig}$a$\footnote{The symbols and sources for Figure 
\ref{Lx_Li_age.fig} are obtained as follows.  GJ 3305 (this paper) 
and CD $-64^\circ$1208 \citep{Zuckerman01} are shown as large open 
stars. For the X-ray data: young PMS stars are from the Orion Nebula 
Cluster \citep[][circles]{Getman05} and the Taurus-Auriga cloud 
\citep{Neuhauser95}, $\sim 10$ Myr PMS stars are from the TW Hya 
Association \citep[][squares]{Webb99}, $\sim 30$ Myr stars are from 
NGC 2547 \citep{Jeffries98}, $\sim 100$ Myr stars are from the 
Pleiades \citep{Stauffer94}, and $>1$ Gyr old stars are from the 
Gliese catalogue \citep{Schmitt04}.  For the lithium data: young PMS 
stars are from the Taurus-Auriga clouds \citep[][circles]{Martin94} 
and $\sigma$ Orionis cluster \citep[][squares]{Zapatero02}, old PMS 
stars are from the OSCA Upper Scorpius subgroup 
\citep[][circles]{Preibisch98} and the TW Hya Association 
\citep[][squares]{Webb99, Sterzik99, Zuckerman01b}, and $\sim 30$ 
Myr stars from the IC 2602 and IC 2391 clusters 
\citep[][circles]{Randich01} and NGC 2547 cluster 
\citep[][squares]{Jeffries03}. \label{Lx_Li.footnote}}.  Due to 
mass-dependencies in pre-main sequence X-ray emission 
\citep{Preibisch05}, we limit the comparison to stars in a limited 
spectral range.  The saturated X-ray level implies an age $\la 100$ 
Myr.  

\subsection{HR diagram \label{hrd.sec}}

In Figure \ref{hrd.fig} we compare the location of 51 Eri and GJ 
3305 in the HR diagram to the predictions of the PMS evolutionary 
tracks of \citet{Siess00}. The error bars indicate the effect of a 
nominal 0.5 sub-type uncertainty in the spectral types of the two 
stars; F0 for 51 Eri and M0.5 for GJ 3305.  The resulting range in 
the stellar luminosities caused by the adoption of the 0.5 sub-type 
uncertainty is encompassed by the plotted size of the points in 
Figure \ref{hrd.fig}.  51 Eri may lie slightly elevated above the 
ZAMS of \citet{Siess00}, although only marginally-so owing to the 
adopted uncertainty and the coarseness of the grid in the 
temperature range appropriate for F-type stars.  At 30 pc distance, 
GJ 3305 is clearly PMS with an inferred age of $13^{+4}_{-3}$ Myr, 
consistent with the age estimate of $12^{+8}_{-4}$ Myr given for the 
BPMG by \citet{Zuckerman01} and the 17 Myr age given for the rich 
LCC OSCA subgroup by \citet{Mamajek02}.  From comparison with the 
models of \citet{Siess00} we obtain $M = 1.6$ M$_{\odot}$ for 51 
Eri, and $M = 0.5$ M$_{\odot}$ for GJ 3305.

\subsection{Multiplicity and dynamical state}

If we assume that the physical separation of 51 Eri/GJ 3305 is close 
to the projected separation of 66\arcsec, or $\simeq 2000$ AU at $d 
= 30$ pc, and that the orbit is circular, then the binary orbital 
period is $\simeq 60,000$ yr and an orbital velocity of GJ 3305 is 
$\approx 0.7$ km\,s$^{-1}$.  This represents a very fragile system; 
any dynamical perturbation of order 1 km\,s$^{-1}$ during the last 
$10-15$ Myr would have been sufficient to disrupt the binary, yet 
the binary system has completed $\sim 200$ orbital periods.  This is 
reminiscent of a similar multiple system: the four low-mass PMS 
stars with $500-1500$ AU orbits around the $3-5$ Myr 
intermediate-mass Herbig Ae star HD 104237 which is also 
kinematically linked to the OSCA \citep{Feigelson03}.  Systems such 
as 51 Eri and HD 104237 must have originated in a dynamically 
quiescent star formation environment. Dynamical simulations of 
stellar clusters by \citet{Kroupa98} indicates that wide binaries 
with total mass around 2 M$_\odot$ and mass ratio around 3:1, as in 
51 Eri/GJ 3305, can survive if the initial star density is not too 
high.  

\subsection{Rotational velocities}

The $v\sin i$ projected rotational velocity of 51 Eri was measured 
to be $71.8 \pm 3.6$ km\,s$^{-1}$ by \citet{Reiners03}, with 
indications from suspected asymmetries and variations in the line 
profiles that the star was spotted.  51 Eri is listed as a candidate 
{\it Hipparcos\,}variable star by \citet{Koen02}, with a period 
determined by Fourier analysis of the {\it Hipparcos\,} photometry 
of $P = 0.65$ d, with a fitted sine wave amplitude of 0.005 mag.  
For a F0 dwarf on or near the main-sequence with characteristic mass 
and radius of $M \approx 1.5$M$_{\odot}$ and $R \approx 1.5$ 
R$_{\odot}$, the $v\sin i$ velocity and proposed period of the star 
can be reconciled if 51 Eri is being observed at a moderate 
inclination angle of $i \sim 45^{\circ}$.

There is no literature value for the projected rotational velocity 
for GJ 3305; however our HET spectra for GJ 3305 show line profiles 
for absorption lines near H$\alpha$ of width exceeding the expected 
instrumental broadening due to the HRS.  From the measured FWHM of 
the profiles we estimate the contribution from all sources of line 
broadening to be $v \sim 10$ km\,s$^{-1}$.  Thus $v\sin i < 10$ 
km\,s$^{-1}$ and for a young $R \approx 0.7$ R$_{\odot}$ early-M 
star the inferred rotation period is $P > 4\sin i$ days, consistent 
with our tentative measured rotation period of $P = 6.1$ days; see 
\S 2.3.  We have not attempted a line profile deconvolution that 
would allow us to accurately determine the projected rotational 
velocity for the star.

\section{Origin of the 51 Eri and GJ 3305 system \label{orig.sec}}

There is little doubt that 51 Eri and its companion constitute a 
physical binary.  Not one of several thousand UCAC2 stars in the 
immediate vicinity share the unusually large proper motions seen in 
these two stars. 

We conclude that both the 1.6 M$_\odot$ 51 Eri and its X-ray 
selected companion 0.5 M$_\odot$ companion are $\simeq 10$ Myr 
outlying members of the OSCA.  They are now seen at an angular 
distance of $\simeq 100^\circ$, and a physical distance of 
$\simeq110$ pc, from the center of the nearest (LCC) rich 
concentration of OSCA stars.  This implies that the 51 Eri system 
was formed from molecular gas with a velocity vector displaced $\sim 
9$ km\,s$^{-1}$ from that of the rich OSCA concentrations. This is 
consistent with dispersions seen over large scales in giant 
molecular cloud complexes which are attributed to turbulence 
\citep{Feigelson96}.  

OSCA outliers include a number of PMS stars with rapid gas accretion 
from infrared-bright ({\it IRAS}-detected) circumstellar disks.  
These include TW Hya and Hen 3-600A in the TWA \citep{Muzerolle00}, 
ECHAJ 0843.3--7905 and RECX 11 in the $\eta$ Cha cluster 
\citep{Lawson04}, and the several Herbig Ae/Be stars noted in Figure 
\ref{OSCA.fig}.  $\beta$ Pic has a prominent reflection disk and is 
accreting cometary-sized bodies \citep{Lagrange00}.  In contrast, 
both 51 Eri and GJ 3305 appear to lack prominent disks.  51 Eri 
displays a $(V-K)$ color of 0.68, consistent with its F0 spectral 
type.  However, the star may have a small $(K-[12])$ excess of $0.2 
\pm 0.1$ when compared to dwarf colors assembled by 
\citet{Kenyon95}, which may indicate the presence of a weak disk.  
GJ 3305 has 2MASS {\it JHK\,} colors consistent with its listed 
spectral type of M0.5; however it has $(V-[{\rm 2MASS}])$ colors 
closer to that of an M2 dwarf which may indicate the presence of a 
weak disk, or a minor mis-classification of its spectral type.  We 
note that GJ 3305 has no known late-M companion; \citet{McCarthy04} 
included GJ 3305 in an unsuccessful search to detect brown dwarfs 
orbiting at $a = 75-300$ AU around a hundred nearby G-, K-, and 
M-type stars.

Sensitive near-infrared coronographic imagery and far-infrared 
photometry with the {\it Spitzer Space Observatory\,} may detect 
faint, evolved disks around one or both stars.  Perhaps most 
exciting is the possibility that, with disks that have largely 
dissipated, protoplanets might be present around these stars.  For 
example, making use of the sub-stellar models of \cite{Baraffe02}, a 
5 M$_J$ planet in a circular face-on orbit with $a = 30$ AU should 
appear as a $K = 15.4$ source at a separation of 1\arcsec\/ from the 
star.  For such a planet surrounding GJ 3305, $\Delta K = 9$ mag; 
this is a demanding but not insurmountable challenge using existing 
technologies.

\section{HD 172555/CD -64$^\circ$1208 and other similar wide binary 
systems}

While it may seem at first unlikely that a weakly bound binary can 
survive the turbulent environment of the OSCA giant molecular cloud, 
a considerable fraction of the identified BPMG members are similar.  
Visual multiple systems in the list of \citet{Zuckerman01} include 
51 Eri/GJ 3305 (F0+M0.5 with projected separation  2200 AU), HD 
155555ABC (G5+K0+M4.5, 400 AU), HD 172555/CD -64$^\circ$1208 (A7+K7, 
2000 AU), HR 7329AB (A0+M7, 200 AU), and HD 199143AB/BD 
-17$^\circ$6128AB \citep[primaries F8+K7, 15000 AU][]{Kaisler04}. 

Of these, the HD 172555/CD -64$^\circ$1208 system is remarkably 
similar to 51 Eri/GJ 3305.  HD 172555 = HR 7012 is a mid-A star 
lying far to the southeast of the main OSCA concentrations in the 
sky (Figure \ref{OSCA.fig}). Its spectral type has been variously 
classified as A2 to A7; \citet{Houk75} give A5 IV--V while 
\citet{Gray89} give A6 IV based on spectral line shapes.  With a 
{\it Hipparcos\,} parallactic distance of $29.2 \pm 0.6$ pc, an 
enormous proper motion oriented southward ($\mu_\alpha, \mu_\delta$) 
= ($32.7, -148.7$) mas~yr$^{-1}$, and a roughly measured radial 
velocity of $+2 \pm 5$ km s$^{-1}$, its closest approach to the OSCA 
was 11 Myr ago when it lay $\sim 29$ pc from the Upper Centaurus 
Lupus (UCL) subgroup. 

This intermediate-mass star has a comoving companion CD 
$-64^\circ$1208 with spectral type M0, $V = 10.4$ and $K = 6.1$.  
Whereas the older literature did not indicate a relationship between 
51 Eri and GJ 3305, various proper motion surveys since 
\citet{Fallon83} showed that HD 172555 and CD $-64^\circ$1208 are 
approximately comoving.  The most accurate proper motion for the M0 
star is from the UCAC2 catalog with ($\mu_\alpha, \mu_\delta$) = 
($30 \pm 14, -153 \pm 14$) mas~yr$^{-1}$ which is consistent with 
the A star.  Only 0.04\% of stars within $\approx 2.5^{\circ}$ of HD 
172555 have UCAC2 motions within $\pm 20$ mas\,yr$^{-1}$ of its 
value.  

A spectrum of CD $-64^\circ$1208 reported by \citet[][footnote to 
Table 1]{Zuckerman01} shows H$\alpha$ in emission with 
$EW$(H$\alpha$) $= 2.2$~\AA\/ and strong lithium in absorption with 
$EW$(Li I $\lambda$6708) $= 0.58$~\AA.  Its X-ray emission is very 
strong, about 1 ct\,s$^{-1}$ in ROSAT proportional counter 
observations \citep{Simon93} with a soft X-ray spectrum, strong 
extreme-ultraviolet emission and possibly the source of a soft X-ray 
flare seen with the early non-imaging $Ginga$ satellite 
\citep{Kreysing95, Thomas98, Forster99}. We have obtained a brief 
$Chandra$ snapshot of the region with the ACIS detector (ObsID = 
5180, 2004 Mar 10; not displayed here).  The mid-A primary is not 
seen with a limit $\log L_x < 26.9$ erg s$^{-1}$ ($0.5-8$ keV) but 
the M0 secondary is detected with $\simeq 0.2$ ct s$^{-1}$.  The 
corresponding luminosity at $d=29$ pc is $\log L_x \simeq 29.5$ erg 
s$^{-1}$.  Similar to GJ 3305, this is near the top of the X-ray 
luminosity function for M dwarfs and indicates a very young age.

These two binary systems, 51 Eri/GJ 3305 and HD 172555/CD 
$-64^\circ$1208, share many properties: a (coincidental) distance of 
30 pc from the Sun, a factor of $3-4$ mass ratio, a projected 
separation of 2000 AU, very strong X-ray emission in the lower mass 
star, compatible very-high velocity proper motions pointed towards 
an origin in the OSCA, and a location of the lower mass star above 
the main sequence assuming it is codistant with the higher mass 
star.  They differ only in their distant locations from each other 
and the OSCA concentrations in the sky, and in the strength of the 
lithium line which is 5 times stronger in CD $-64^\circ$1208 than GJ 
3305.  \citet{Masciadri05} have conducted a high-resolution 
near-infrared imaging search for planetary companions around CD 
$-64^\circ$1208; no gaseous companion with mass $>5$ M$_J$ was found 
with orbital radius $>15$ AU.
  
\section{Discussion}

The evidence is clear that 51 Eri and GJ 3305 together are a young, 
wide binary system at $d = 30$ pc. The distance to 51 Eri is 
directly measured by parallax, and has a high-velocity space motion 
that is convergent with the OSCA.  GJ 3305 shares its unusual proper 
motion, and is thus very unlikely ($P < 0.0002$) to be an unrelated 
star.  Assuming a 30 pc distance, GJ 3305 lies on the $\simeq 13$ 
Myr isochrone on the H-R diagram.  It exhibits moderately strong 
chromospheric emission and extreme X-ray emission with powerful 
X-ray flares.  Its photometric variability is probably due to 
starspots modulated with a 6-day period. Its spectrum shows an 
abundance of lithium intermediate between the primordial levels seen 
in young T Tauri stars and the fully depleted main sequence stars.  
From this confluence of evidence, we conclude with confidence that 
both 51 Eri and GJ 3305 are OSCA outliers with ages around $10-15$ 
Myr similar to the UCL and LCC OSCA subgroups. 

The main anomaly is the relatively weak Li I $\lambda$6708 line 
compared to other OSCA stars (Figure \ref{Lx_Li_age.fig}$b$).  
However, it is well-known that ZAMS stars exhibit a wide range of 
lithium levels and there is growing evidence that a small fraction 
of M stars undergo rapid lithium depletion while on their pre-main 
sequence tracks.  In addition to the BPMG stars discussed in \S 
\ref{het.sec}, \citet{Song02} report a M4+M4.5 BPMG binary where the 
two components show an order of magnitude difference in lithium line 
strength ($EW = 300$ m\AA\/ and $EW < 30$ m\AA). A spectroscopic 
survey for lithium in $\sim 10$ Myr Orion Nebula Cluster stars shows 
a few lithium-depleted stars contemporaneous with the dominant 
population of undepleted stars \citep{Palla05}.  Conversely, a 
survey of the 35 Myr NGC 2547 cluster \citep{Jeffries03} reveals 
several stars with intermediate levels of lithium contemporaneous 
with the dominant population of lithium-depleted stars.  Different 
theoretical models of M0 stellar interiors give very different 
predictions of lithium depletion in the $10-20$ Myr age range (see 
Figure 9 in Jeffries et al.). Detailed modeling of solar mass PMS 
stars shows that slow or rapid lithium depletion in $10-20$ Myr can 
readily be explained by small changes in convection zone properties 
\citep{Piau02}.  It is thus reasonable to conclude that GJ 3305 is 
an OSCA member that experienced rapid depletion of lithium.  

51 Eri and GJ 3305 lack significant $K$-band excesses and thus lack 
massive inner protoplanetary disks.  Stars devoid of inner disks at 
$t \simeq 10$ Myr are known in other OSCA outlying groups, e.g. in 
the TWA \citep{Uchida04} and the $\eta$ Cha cluster \citep{Lyo03}, 
and have been inferred from the wide range of rotation rates 
measured in the Pleiades and other ZAMS clusters.  Such stars are 
excellent candidates for searches for outer disks at mid-infrared 
wavelengths using the {\it Spitzer Space Telescope}, and high 
spatial resolution surveys for Jovian-mass protoplanets.

\begin{figure}
\centering
\includegraphics[height=1.0\textwidth,angle=-90]{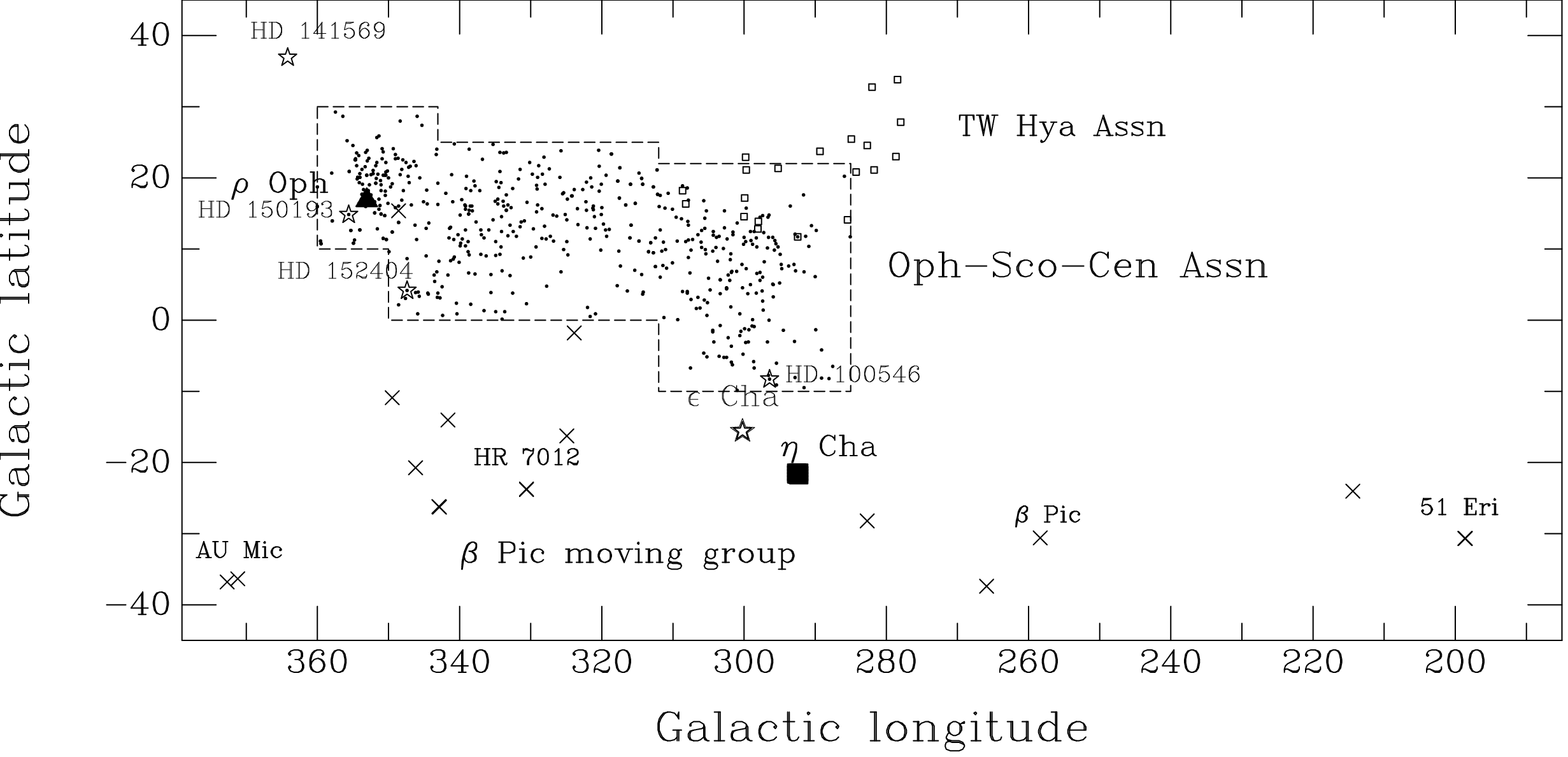}
\caption{Diagram of $\sim 1/3$ of the celestial sphere in Galactic 
coordinates showing nearby stars and PMS groups kinematically 
associated with the Oph-Sco-Cen Association (OSCA).  References and 
further discussion are given in Feigelson et al. (2003). 
\label{OSCA.fig}}
\end{figure}

\begin{figure}
\centering
\includegraphics[width=3.5in]{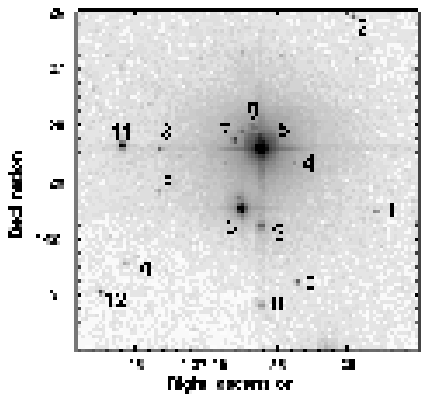}
\includegraphics[width=3.5in]{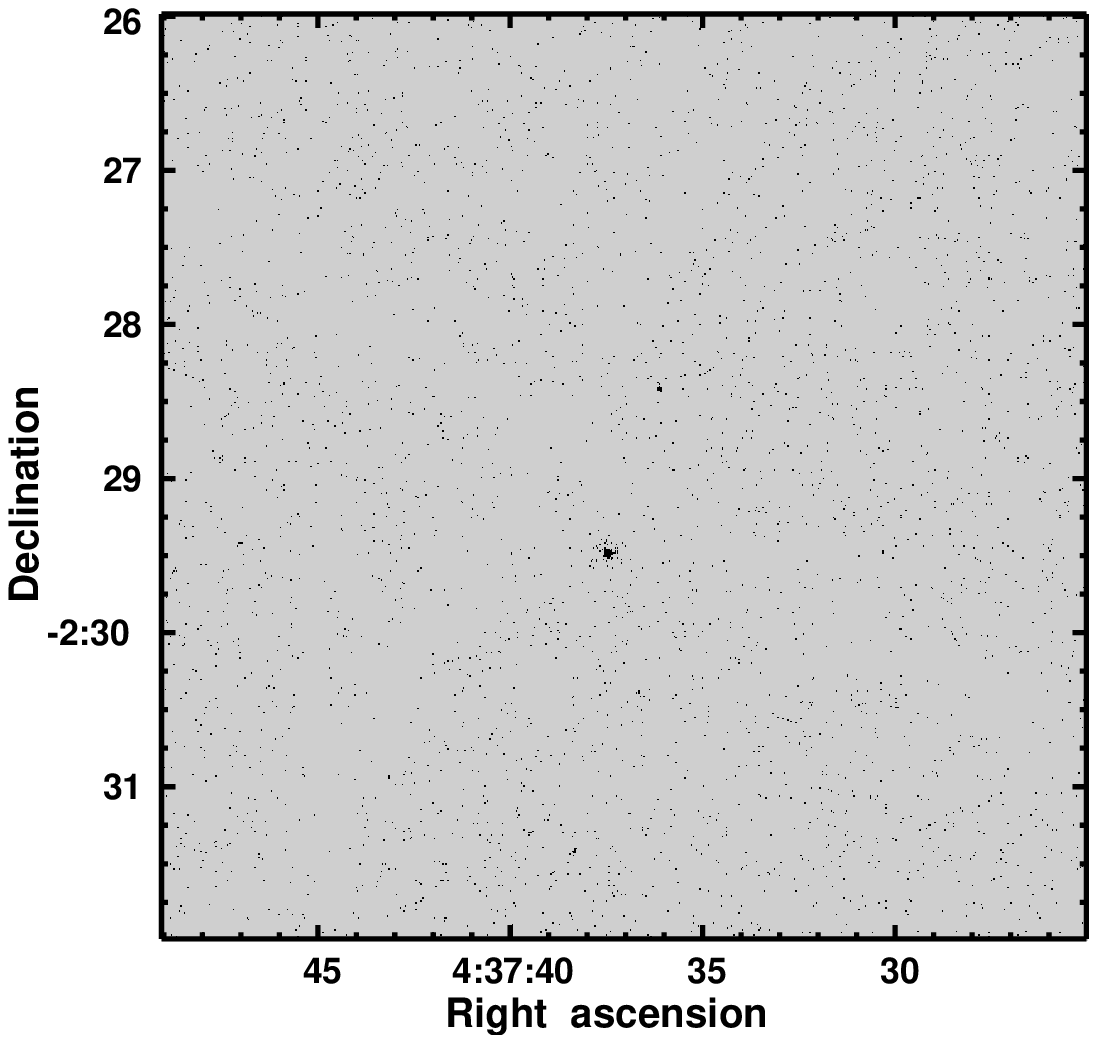}
\caption{(Top) A 6\arcmin$\times$6\arcmin\/ region around 51 Eri 
from the 2MASS $J$-band survey.  The numbered stars are listed in 
Table \ref{vicinity.tab}, and `g' refers to instrumental glint and 
ghost images.  Star 5 is 51 Eri and star 6 is GJ 3305. (Bottom) The 
same region from the {\it Chandra\,} ACIS image showing 51 Eri 
(faint, slightly north of center) and GJ 3305 (bright, slightly 
south of center). \label{images.fig}}
\end{figure}

\begin{figure}
\centering
\includegraphics[width=0.6\textwidth,angle=-90]{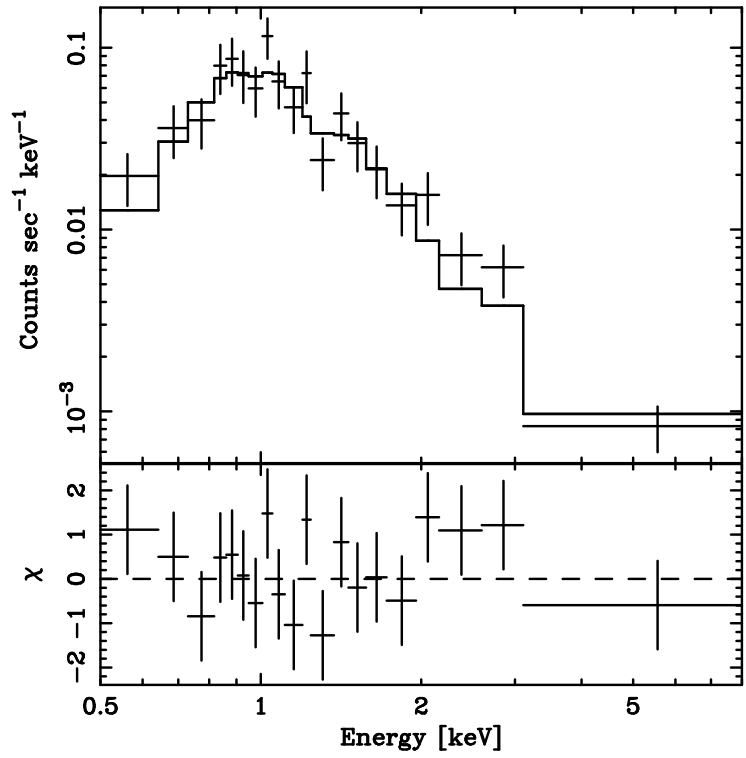}
\caption{{\it Chandra\,} ACIS spectrum of GJ 3305. 
\label{xspec.fig}}
\end{figure}

\begin{figure}
\centering
\includegraphics[width=0.9\textwidth]{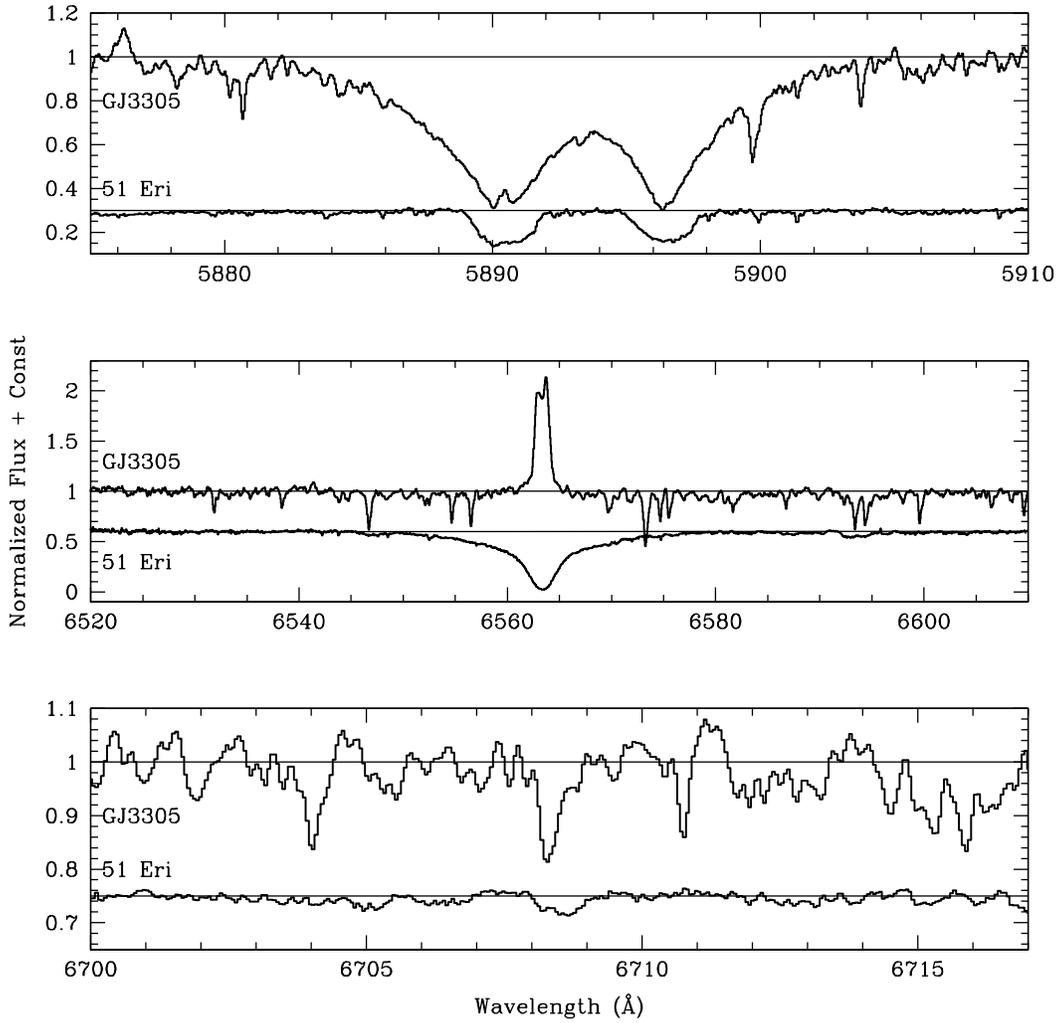}
\caption{High-resolution echelle spectra from the Hobby-Eberly 
Telescope of 51 Eri and GJ 3305: (top) Na D region, (middle) 
H$\alpha$ region, and (bottom) Li\,6708\AA\, region. The 51 Eri 
spectra are offset from unity normalization for clarity.  
\label{het.fig}}
\end{figure}

\begin{figure}
\centering
\includegraphics[width=0.6\textwidth]{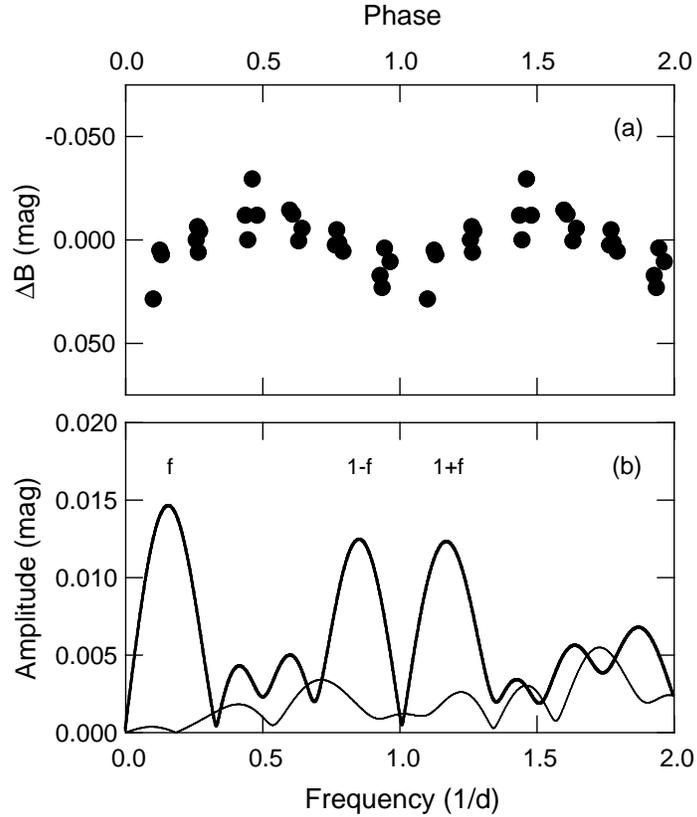}
\caption{(Top) SAAO $B$-band observations of GJ 3305 phased to a 
periodicity of $f = 0.164$ d$^{-1}$ ($P = 6.1$ days).  (Bottom) 
Amplitude spectra of the original $B$-band dataset (upper bold line) 
and the pre-whitened dataset following removal of the $f = 0.164$ 
d$^{-1}$ periodicity (lower thin line).  The $\pm 1$ d$^{-1}$ 
aliases of the periodicity resulting from observations being made at 
a single observing site are identified.   \label{phase.fig}}
\end{figure}

\begin{figure}
\centering
\includegraphics[width=0.5\textwidth,angle=-90.]{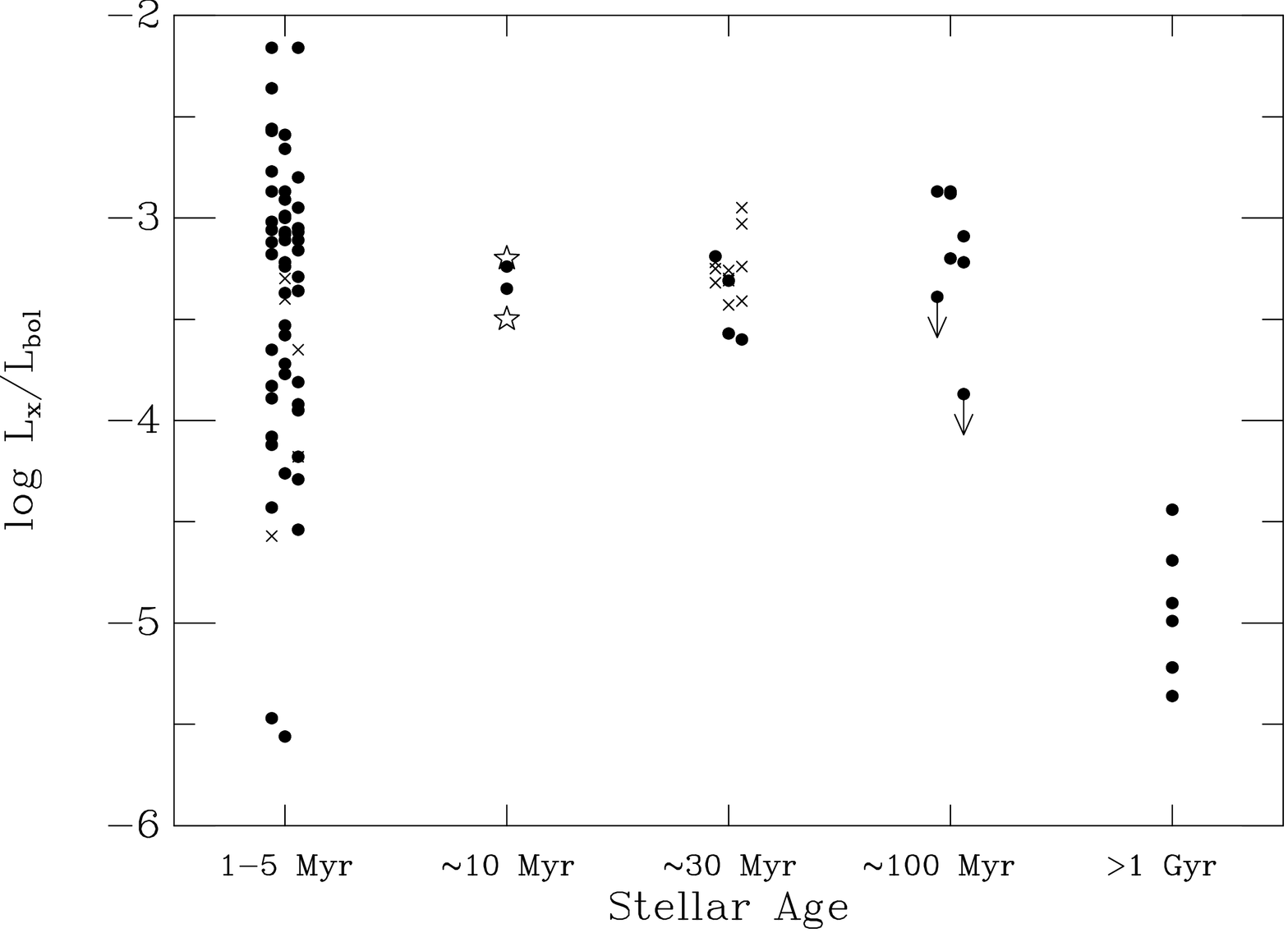}
\includegraphics[width=0.5\textwidth,angle=-90.]{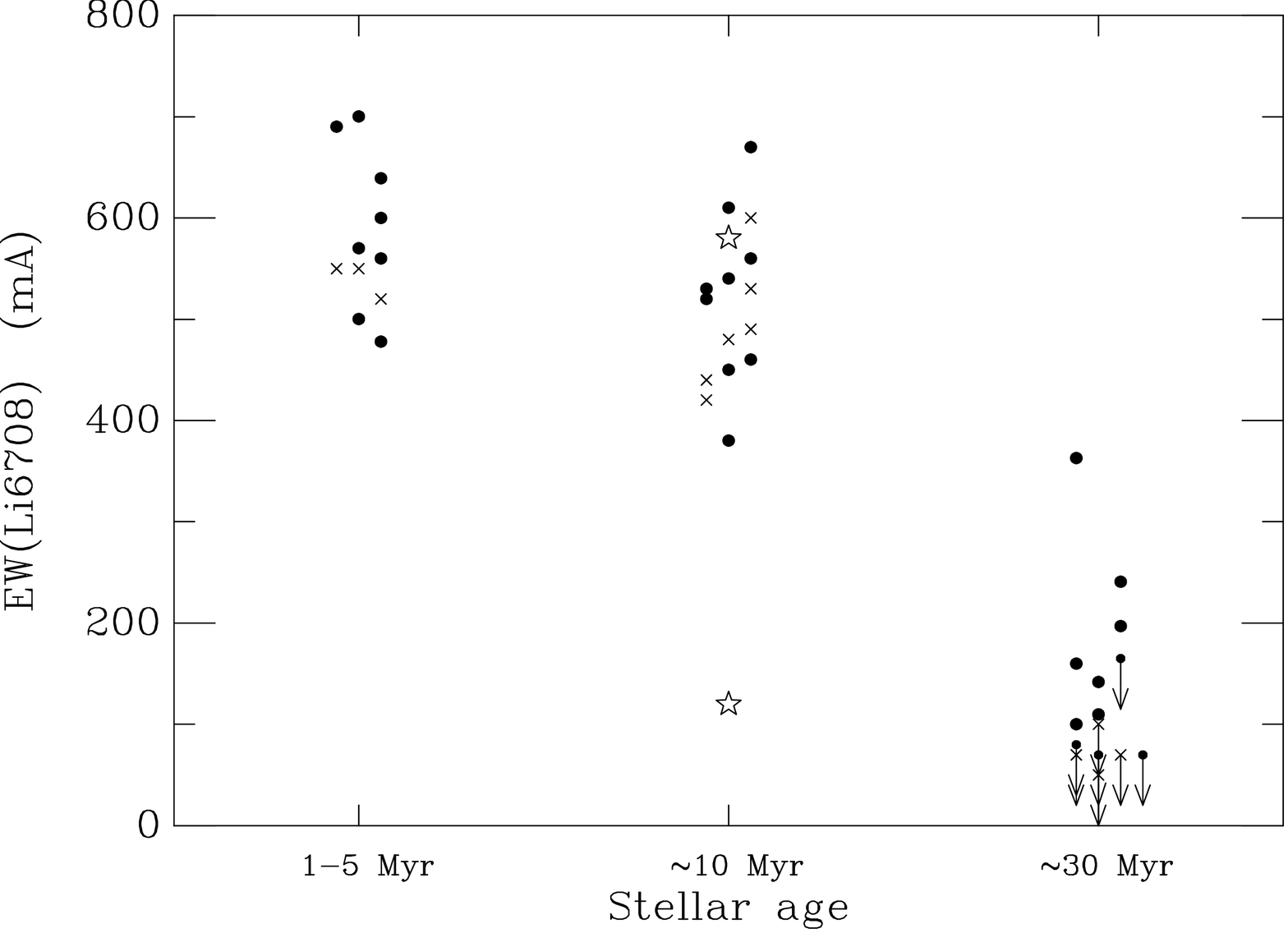}
\caption{Distributions of (top) X-ray emissivity $\log L_x/L_{bol}$ 
and (bottom) lithium 6708\AA\/ line equivalent width as a function 
of stellar age for stars with spectral types in the narrow range 
M0.0-M1.0.  GJ 3305 and CD -64$^\circ$1208 are shown as large open 
stars.  Other symbols and details of the plotted values are given in 
footnote \ref{Lx_Li.footnote}  \label{Lx_Li_age.fig}}

\end{figure}

\begin{figure}
\centering
\includegraphics[width=0.9\textwidth]{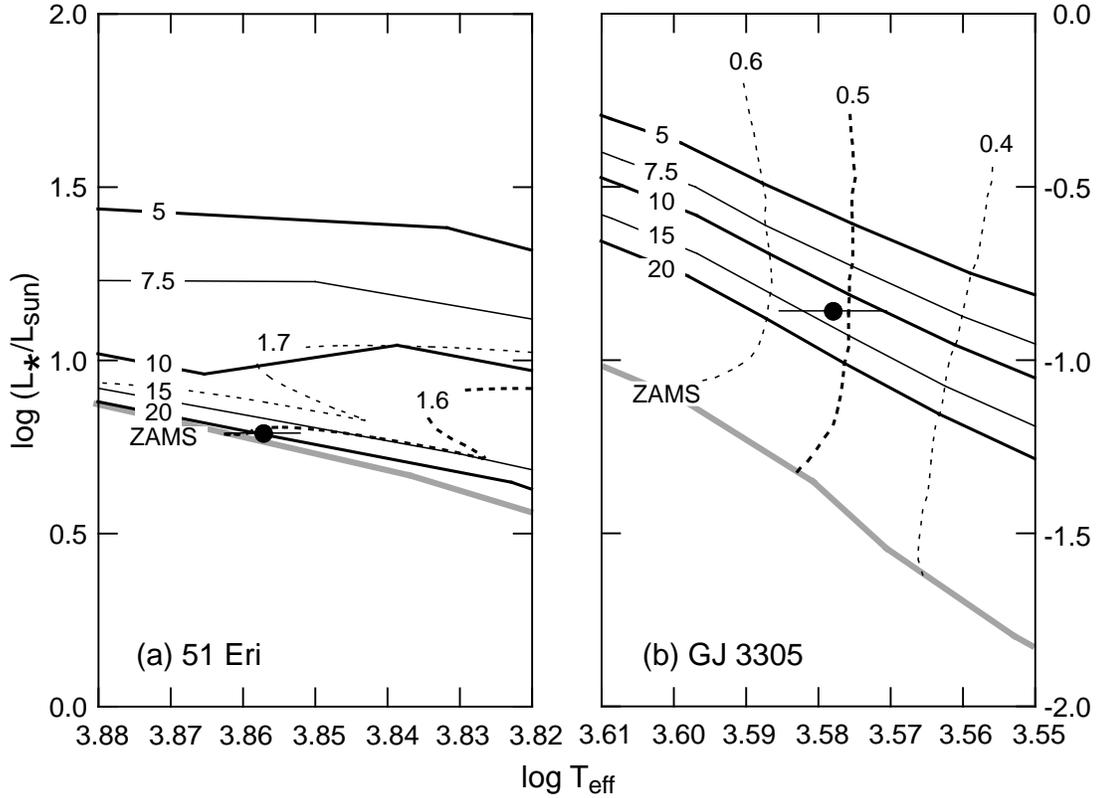}
\caption{HR diagram location for (left) 51 Eri and (right) GJ 3305, 
compared to the PMS grid of Siess et al. (2000). Error bars for 51 
Eri and GJ 3305 assume a nominal $\pm 0.5$ sub-type in the spectral 
types for these stars of F0 and M0.5, respectively. Dwarf 
temperature and bolometric correction sequences have been assumed. 
Isochrones are in units of Myr; isomass lines in units of 
M$_{\odot}$. For GJ 3305, we adopt an age of 13$^{+4}_{-3}$ Myr. 
\label{hrd.fig}}
\end{figure}

Kinematical evidence indicates the binary system is an outlier of 
the OSCA, similar to members of the BPMG, TWA, $\eta$ Cha cluster, 
and others shown in Figure \ref{OSCA.fig}. The 51 Eri system is the 
most distant member yet found, lying $100^{\circ}$ in the sky and 
110 pc in space from the rich LCC OSCA subgroup with age $\sim 17$ 
Myr \citep{Mamajek02}. This can happen if the molecular cloudlet 
from which the 51 Eri system was formed was imparted a velocity of 
$\sim 9$ km\,s$^{-1}$ with respect to the more massive molecular 
cloud that formed the LCC subgroup.  Such velocity differences are 
commonly seen in giant molecular cloud complexes that give rise to 
OB associations and can be imparted to disperse young stellar 
systems \citep{Feigelson96}.  Yet, the relative velocity of the 51 
Eri and GJ 3305 components could not have remained bound if they 
experienced a relative motion greater than 1 km\,s$^{-1}$.  The HD 
104237 quintet \citep{Feigelson03} and the HD 172555/CD 
$-64^\circ$1208 binary are similarly dynamically fragile multiple 
systems among OSCA outliers.  

The 51 Eri/GJ 3305 binary system is thus a valuable tool in several 
respects.  It validates the use of kinematic extrapolations of 
motions across the sky to find OSCA outliers.  It supports arguments  
for distributed star formation in turbulent giant molecular cloud 
complexes producing small stellar groups which disperse far from the 
rich OB associations.  GJ 3305 can be studied as an example of rapid 
lithium depletion in PMS stars.  Finally, as one of the closest 
stellar systems with an age comparable to the later phases of planet 
formation, both 51 Eri and GJ 3305 represent excellent laboratories 
for the search for older protoplanetary disks or recently formed 
protoplanets. 

\acknowledgements  

This work was supported by NASA contract AR5-6001X (Garmire, PI).  
WAL acknowledges support from a UNSW@ADFA Special Research Grant.  
We thank L. W. Ramsey (Penn State) for rapid access to the 
Hobby-Eberly Telescope, L. A. Crause (SAAO), Charles Poteet (WKU), 
John Cybulski and Ethan Jordan (Penn State) for assistance.  The  
referee, Alex Brown (Colorado) provided very helpful commentary. The 
effort benefited from on-line data resources from the Centre des 
Donn\'ees Stellaires (Strasbourg), Astrophysics Data System 
(Smithsonian Astrophysical Obs.), High Energy Astrophysics Science 
Archive Research Center (NASA-GSFC), and the Infrared Space Archive 
(NASA-IPAC).

\newpage

\clearpage
\newpage

\clearpage
\newpage

\clearpage
\newpage

\clearpage
\newpage

\clearpage
\newpage

\clearpage
\newpage

\clearpage
\newpage

\begin{deluxetable}{rrrccrrrr}
\centering \tablecolumns{9} \tabletypesize{\small} \tablewidth{0pt}

\tablecaption{Stars in the vicinity of 51 Eri \label{vicinity.tab}}

\tablehead{

\colhead{\#} & \colhead{R.A.} & \colhead{Dec.} & \colhead{UCAC \#} & 
\colhead{Other catalogs} & \colhead{UC} & \colhead{$J$}
& \colhead{$\mu_\alpha$} & \colhead{$\mu_\delta$} \\

\colhead{(1)} & \colhead{(2)} & \colhead{(3)} & \colhead{(4)} & 
\colhead{(5)} & \colhead{(6)} & \colhead{(7)} & \colhead{(8)} & 
\colhead{(9)}  }

\startdata
  1 & 28.0 & 29 31 & 30947507 & CMC, USNO                      & 15.0 & 13.5 & 12.3$\pm$7.5 &  -6.8$\pm$7.5 \\
  2 & 29.6 & 26 04 & 30947509 & GSC                            & 14.3 & 13.3 &  5.3$\pm$5.4 & -10.6$\pm$5.4 \\
  3 & 33.5 & 30 36 & 30794265 & CMC, USNO                      & 14.1 & 13.1 &  5.6$\pm$7.6 &  -5.2$\pm$7.6 \\
  4 & 33.8 & 28 39 & 30947515 & APM                            & 15.5 & 14.6 &  7.4$\pm$7.5 &  -1.4$\pm$7.5 \\
  {\bf 5} & {\bf 36.1} & {\bf 28 25} & {\bf \nodata}  & {\bf 51 Eri, SD, WDS, CCDM, \ldots} & {\bf  5.2} & {\bf  4.7} & {\bf 43.3$\pm$0.8} & {\bf -64.3$\pm$0.6} \\
  {\bf 6} & {\bf 37.5} & {\bf 29 28} & {\bf 30947521} & {\bf CMC, GJ, StKM, RBS, \ldots}    & {\bf 10.0} & {\bf  7.3} & {\bf 46.1$\pm$2.8} & {\bf -64.8$\pm$3.0} \\
  7 & 38.0 & 28 16 & 30947523 & WDS, CCDM, HIC                 & 11.8 & 11.3 &  0.6$\pm$1.4 &  20.2$\pm$1.4 \\
  8 & 43.3 & 28 25 & 30947531 & CMC                            & 14.2 & 13.3 & 12.9$\pm$7.5 &  -3.2$\pm$7.6 \\
  9 & 43.4 & 29 11 & 30947532 & CMC                            & 15.9 & 14.4 &  6.3$\pm$7.5 &  -2.8$\pm$7.5 \\
 10 & 45.6 & 30 26 & 30794283 & \nodata                        & 14.6 & 13.6 & -5.9$\pm$7.5 &  -9.5$\pm$7.5 \\
 11 & 45.9 & 28 22 & 30947535 & CMC, GSC, SD, Tycho, \ldots   & 10.9 & 10.3 &  7.8$\pm$2.9 &  -6.0$\pm$3.3 \\
 12 & 47.4 & 30 57 & 30794286 & \nodata                        & 13.1 & 12.1 & -4.8$\pm$7.5 &  -8.7$\pm$7.5 \\
\enddata

\tablecomments{Table columns:
Col 1. Running star number from Figure 2a \\
Col 2-3. Right ascension and declination, J2000 \\
Col 4. UCAC2 star number \citep{Zacharias03} \\
Col 5. APM = Automated Plate Measurement catalog APMCAT-POSS1-1.0 
\citep{Irwin92}; CCDM = Catalogue of the Components of Double and 
Multiple Stars \citep{Dommanget02}; CMC = CMC12 (Version 1.0) 
catalog \citep{Evans02}; GJ = Catalog of Nearby Stars, Preliminary 
3rd version \citep{Gliese95}; GSC2 = Guide Star Catalog (version 
2.2) from the Palomar Schmidt telescope; RBS = ROSAT Bright Survey 
\citep{Voges99}; SD = Southern Durchmusterung \citep{Schonfeld1886}; 
StKM = \citep{Stephenson86}; Tycho = Tycho 2 catalogue 
\citep{Hog00}; USNO = USNO-B1.0 catalog \citep{Monet03}; WDS = 
Washington Visual Double Star Catalog
\citep{Worley97} \\
Col 6. UCAC magnitude (between $V$ and $R$ band, $\pm$0.3\arcsec\/
accuracy), except for 51 Eri ($V$ magnitude) \\
Col 7. $J$ band magnitude from 2MASS All-Sky Point Source Catalog 
Col 8-9.  Proper motion from UCAC2 \citep{Zacharias03}, except for 
51 Eri from {\it Hipparcos}. }

\end{deluxetable}

\newpage

\begin{deluxetable}{lcc}
\centering \tablecolumns{3} \tabletypesize{\normalsize} 
\tablewidth{0pt}

\tablecaption{X-ray properties of 51 Eri and GJ 3305 
\label{xray.tab}}

\tablehead{\colhead{Property} & \colhead{51 Eri} & \colhead{GJ 3305} 
}

\startdata
CXOU        & 043736.12-022824.7    & 043737.46-022928.3 \\
RA          & $04^{\rm h}37^{\rm m}36.12^{\rm s}$ & 
          $04^{\rm h}37^{\rm m}37.46^{\rm s}$ \\
Dec         & $-02^{\circ}$ 28\arcmin 24.7\arcsec &
              $-02^{\circ}$ 29\arcmin 28.3\arcsec \\
Extr counts & 41                    & 222 \\
Soft counts & 41                    & 182 \\
kT          & 0.2                   &  0.6 and 2.8 \\
Flux        & $1.3 \times 10^{-13}$ & $1.6 \times 10^{-12}$ \\
Luminosity  & $1.4 \times 10^{28}$  & $1.7 \times 10^{29}$
\enddata
\end{deluxetable}

\end{document}